\documentclass[twocolumn,twoside]{IEEEtran}

\ifCLASSINFOpdf

\else

\fi

\ifCLASSOPTIONcompsoc
\usepackage[caption=false, font=normalsize, labelfont=sf, textfont=sf]{subfig}
\else
\usepackage[caption=false, font=normalsize]{subfig}
\fi
\usepackage{lipsum}%
\usepackage[dvipsnames]{xcolor}

\usepackage{balance}
\usepackage{multicol}   
\usepackage{cite}
\usepackage{gensymb}
\usepackage{multirow}
\usepackage{graphics}
\usepackage{epsfig}
\usepackage{graphicx}
\usepackage{epstopdf}
\usepackage{textcomp}
\usepackage{amsmath}
\usepackage{mathtools}
\usepackage{filecontents}
\usepackage{lipsum,color}
\usepackage{amssymb}
\usepackage{float}
\usepackage{comment}

\usepackage{amsthm}

\usepackage{times} 
\usepackage{amsthm}  

\usepackage{amsfonts}

\theoremstyle{break}

\begin{document}
\title{Physical Limits of Entanglement-Based Quantum Key Distribution over Long-Distance Satellite Links}

\author{Mohammad~Taghi~Dabiri, ~Mazen~Hasna,~{\it Senior Member,~IEEE}, \\ ~Saif~Al-Kuwari,~{\it Senior Member,~IEEE}, ~and~Khalid~Qaraqe,~{\it Senior Member,~IEEE}
	\thanks{Mohammad Taghi Dabiri and Mazen Hasna are with the Department of Electrical Engineering, Qatar University, Doha, Qatar (e-mail: m.dabiri@qu.edu.qa; hasna@qu.edu.qa).}
	\thanks{Saif Al-Kuwari is with the Qatar Center for Quantum Computing, College of Science and Engineering, Hamad Bin Khalifa University, Doha, Qatar. email: (smalkuwari@hbku.edu.qa).}
	\thanks{Khalid Qaraqe is with the College of Science and Engineering, Hamad Bin Khalifa University, Doha, Qatar (E-mail: kqaraqe@hbku.edu.qa).}
	\thanks{This publication was made possible by NPRP14C-0909-210008 from the Qatar Research, Development and Innovation (QRDI) Fund (a member of  Qatar Foundation). } 
}

\maketitle
\begin{abstract}
	Entanglement-based quantum key distribution (QKD) protocols, such as E91 and BBM92, offer strong information-theoretic security and are naturally suited for satellite-to-satellite QKD (SatQKD) links. However, implementing these protocols over long-distance inter-satellite free-space optical (FSO) channels poses critical physical-layer challenges that are not addressed in the existing literature. In particular, photon losses due to beam divergence, pointing errors, and background noise can severely degrade the key generation rate and quantum bit error rate (QBER), especially under narrow receiver field-of-view (FoV) constraints.
This paper presents a comprehensive performance analysis of entanglement-based inter-satellite QKD, focusing on photon-level modeling and the impact of practical impairments. We develop analytical expressions for signal detection probabilities, background photon influence, multi-pair emissions, and QBER, incorporating key parameters such as link distance, transmitter tracking jitter, receiver misalignment, and photon pair generation rate. Simulation results reveal the nonlinear sensitivity of system performance to tracking error and FoV limitations, and highlight optimal parameter regimes that jointly maximize secret key rate while maintaining QBER below acceptable thresholds. The proposed model provides actionable design insights for reliable and efficient deployment of entanglement-based SatQKD systems.
\end{abstract}

\begin{IEEEkeywords}
	Quantum key distribution (QKD), satellite communication, entanglement, free-space optics (FSO), field of view (FoV).
\end{IEEEkeywords}

\IEEEpeerreviewmaketitle


\section{Introduction}
In the evolving landscape of secure communication, quantum-based communication stands out as a revolutionary emerging technology offering unconditional security. Unlike classical encryption, which relies on mathematical algorithms vulnerable to future supercomputers, quantum communication derives its strength from the fundamental laws of quantum physics. This means that even with unlimited computational power, an adversary cannot break the security without being detected. This is achieved through quantum key distribution (QkD) protocols, such as BB84 \cite{BENNETT20147}, E91 \cite{Ekert91}, and BBM92 \cite{bennett1992quantum}. 

While QKD has seen significant experimental progress in laboratory and short-distance fiber settings, scaling these systems to long-distance, real-world deployments remains challenging. Early experiments, typically conducted over stable, terrestrial platforms, helped validate foundational quantum principles and calibrate devices such as single-photon sources and detectors. However, moving beyond controlled indoor or metro-scale fiber networks introduces severe technical barriers, including high photon loss, environmental disturbances, and tight requirements on synchronization and pointing accuracy. These limitations are particularly pronounced in free-space and satellite-based scenarios, where the lack of stable infrastructure, large propagation distances, and dynamic alignment errors pose serious constraints on reliability and key throughput. Consequently, the practical realization of scalable QKD systems, especially over space links, is still an open and demanding research problem.

In recent years, substantial efforts have been directed toward extending QKD beyond laboratory and fiber-based environments into longer-range free-space settings. Experimental demonstrations have successfully realized QKD over urban free-space links ranging from 145\,m~\cite{avesani2021full}, to 400\,m~\cite{zhang2022device}, and up to 1.7\,km under daylight conditions~\cite{krvzivc2023towards}. These milestones have validated the feasibility of operating QKD under realistic conditions outside the lab. In parallel, theoretical analyses have highlighted the scalability potential of free-space quantum links for secure and high-rate key distribution~\cite{pirandola2021limits,pan2023free,davidson2024airqkd}.

Building on these foundational studies, researchers have explored the integration of QKD into emerging mobile and infrastructure-constrained environments. Examples include vehicular and transportation networks—such as V2X, V2I, and high-speed rail systems—which demand secure end-to-end communication in dynamic scenarios~\cite{yuan2023analysis,stavdas2024quantum,al2024using}. Additional studies have addressed coexistence challenges between classical and quantum signals over shared optical channels, particularly in the presence of turbulence, pointing errors, and modulation constraints~\cite{alshaer2021hybrid,mantey2025coexistence}. To improve adaptability and robustness, intelligent reconfigurable surfaces (RIS) have recently been proposed as a means to reestablish line-of-sight paths and enhance multi-user QKD under harsh propagation conditions~\cite{chehimi2025reconfigurable,kisseleff2023trusted}. Moreover, airborne quantum networks employing UAVs and high-altitude platforms (HAPs) have been investigated as agile and scalable platforms for QKD, offering flexible topologies in the lower atmosphere~\cite{alshaer2022performance,alshaer2021reliability,alshaer2024exploring}.

These advancements have set the stage for the next frontier: space-based quantum communication. In particular, satellite-to-satellite QKD has emerged as a compelling approach for establishing secure long-distance entanglement links, free from terrestrial infrastructure and atmospheric limitations. Several recent studies have focused on architectural aspects of global quantum networks, with~\cite{de2023satellite} analyzing orbital feasibility,~\cite{gundougan2021proposal} proposing quantum repeater schemes for space links, and~\cite{sun2025quantum} comparing switching strategies for multi-user satellite-based networks. In addition,~\cite{vazquez2023quantum} demonstrates that quantum detection methods can outperform classical ones in pure-loss space channels, even under practical modulation formats such as OOK and BPSK.

While these contributions offer valuable insights into system design and network-level optimization, they often rely on abstracted models that overlook key physical-layer challenges. Among the most critical of these is the impact of tracking error on entanglement-based inter-satellite QKD. Over long propagation distances, even small angular deviations can lead to severe photon misalignment, drastically reducing photon reception probabilities and degrading the overall key rate~\cite{pan2023free,miller2023vector}. 

To better understand these limitations, recent studies have begun modeling the effects of realistic channel impairments in satellite-based QKD (SatQKD). For example,~\cite{sidhu2023finite} investigates finite-key effects and hardware constraints in satellite deployments;~\cite{yang2025performance} analyzes the key rate of satellite-to-ground QKD under beam misalignment; and~\cite{orsucci2024assessment} studies architectural trade-offs involving orbital geometry and link direction. 
However, these studies, while insightful, rarely offer detailed physical-layer modeling for QKD protocols, especially under the practical constraints of long-distance satellite-to-satellite (Sat-Sat) communication. In particular, existing analyses often abstract away the compounded effects of beam divergence, tracking error, and background noise on entanglement distribution and secure key generation.

Among the many QKD protocols, entanglement-based schemes like BBM92 and E91 stand out due to their inherent device-independence, strong security guarantees, and natural suitability for symmetric space-based deployment.
The entanglement-based protocols are particularly well-suited for Sat-Sat communication, where a central entangled photon source can distribute qubits to two distant satellites, enabling global-scale secure key sharing without requiring trusted nodes. However, despite its theoretical appeal, implementing entanglement-based protocols over long-distance satellite links faces two fundamental challenges:
\begin{itemize}
	\item \textbf{Severely limited key generation rate:} Due to beam spreading and angular misalignment, the probability of photon detection at both ends becomes extremely small over long distances, especially in the presence of pointing errors.
	
	\item \textbf{Elevated quantum bit error rate (QBER):} Even when photons are successfully detected, the presence of uncorrelated background photons—amplified by wide receiver Field of View (FoV) or detector noise—leads to high logical error rates that compromise key security.
\end{itemize}
These issues necessitate a deeper physical-layer analysis to identify critical performance bottlenecks and enable practical parameter tuning for reliable entanglement-based satellite QKD, an aspect that, to the best of our knowledge, has not been rigorously investigated in the existing literature.

To address these gaps, this work presents a comprehensive performance analysis of long-distance entanglement-based QKD over inter-satellite links using the entanglement-based protocol. This will provide an important theoretical benchmark for the performance of entanglement-based protocols, which can later be extended to include other protocols. We develop a full-stack photon-level model that captures major physical-layer phenomena, including optical beam propagation, tracking misalignment, background photon interference, receiver FoV filtering, and entangled photon pair emission statistics. Building on this model, we investigate the interplay between link length, transmitter pointing jitter, receiver alignment fluctuations, and background noise conditions. We quantify their effects on both the secret key rate and QBER, and identify fundamental limits and trade-offs in system design. Our framework provides a pathway for optimizing system parameters to enhance key throughput while maintaining QBER below security thresholds.
The key contributions of this paper are summarized as follows:
\begin{enumerate}
	\item We develop a detailed physical-layer model for entanglement-based inter-satellite QKD, incorporating beam divergence, tracking error, receiver aperture coupling, and detection constraints.
	
	\item We derive analytical expressions for the signal photon reception probability, background noise impact, and conditional error rates as functions of system parameters such as link distance, FoV angle, and angular jitter variance.
	
	\item We model the effect of multi-photon events and Poisson-distributed background interference on QBER, explicitly characterizing the dominant error mechanisms in long-distance entanglement-based QKD setups.
	
	\item We investigate the influence of entangled photon pair generation rate on the trade-off between secret key throughput and error rate, and identify optimal operating points that maximize performance under constraints.
	
	\item We provide simulation results under realistic inter-satellite conditions, demonstrating the importance of jointly tuning FoV, photon generation rate, and tracking precision to achieve high-rate, low-QBER operation.
\end{enumerate}

\begin{table*}[!t]
	\centering
	\caption{Summary of Main Notations}
	\label{tab:notations}
	\begin{tabular}{|c|l|}
		\hline
		\textbf{Symbol} & \textbf{Description} \\ \hline
		\( q \) & Channel index (\( A \) for Alice, \( B \) for Bob) \\ \hline
		\( n_t \) & Number of generated entangled photon pairs per time slot \\ \hline
		\( \mu_t \) & Mean number of entangled photon pairs per time slot \\ \hline
		\( n_{r,q} \) & Number of received signal photons at receiver \( q \) \\ \hline
		\( n_{b,q} \) & Number of background photons at receiver \( q \) \\ \hline
		\( \mu_{b,q} \) & Mean number of background photons per time slot at receiver \( q \) \\ \hline
		\( \Phi_{b,q} \) & Background photon flux density at receiver \( q \) \\ \hline
		\( \Omega_{\mathrm{FoV},q} \) & Receiver field-of-view solid angle \\ \hline
		\( \theta_{\mathrm{FoV},q} \) & Receiver field-of-view angle \\ \hline
		\( \eta_q \) & Overall transmittance of the channel \( q \) \\ \hline
		\( \eta_{p,q} \) & Aperture coupling efficiency for channel \( q \) \\ \hline
		\( \eta_{\mathrm{det},q} \) & Detector quantum efficiency at receiver \( q \) \\ \hline
		\( \eta_{r,q} \) & Mean number of detected signal photons at receiver \( q \) \\ \hline
		\( w_{z,q} \) & Beam width at receiver plane for channel \( q \) \\ \hline
		\( a_q \) & Receiver aperture radius for channel \( q \) \\ \hline
		\( Z_{L,q} \) & Propagation distance for channel \( q \) \\ \hline
		\( \sigma_{\theta,t,q}^2 \) & Transmitter angular tracking error variance for channel \( q \) \\ \hline
		\( c_q \) & Scaling factor \( c_q = \mu_t \eta_{\mathrm{det},q} \) \\ \hline
		\( \gamma_q \) & Parameter \( \gamma_q = \frac{w_{z,q}^2}{4Z_{L,q}^2\sigma_{\theta,t,q}^2} \) \\ \hline
		\( P_{\mathrm{click},q} \) & Probability of detecting at least one photon at receiver \( q \) \\ \hline
		\( P_{\mathrm{err}}(\eta_{r,A}, \eta_{r,B}) \) & Conditional expected number of erroneous bits \\ \hline
		\( P_{\mathrm{click}}(\eta_{r,A}, \eta_{r,B}) \) & Conditional expected number of sifted key bits \\ \hline
		\(\text{QBER}(\eta_{r,A}, \eta_{r,B})\) & Conditional quantum bit error rate \\ \hline
	\end{tabular}
\end{table*}

\begin{figure}
	\begin{center}
		\includegraphics[width=3.3 in]{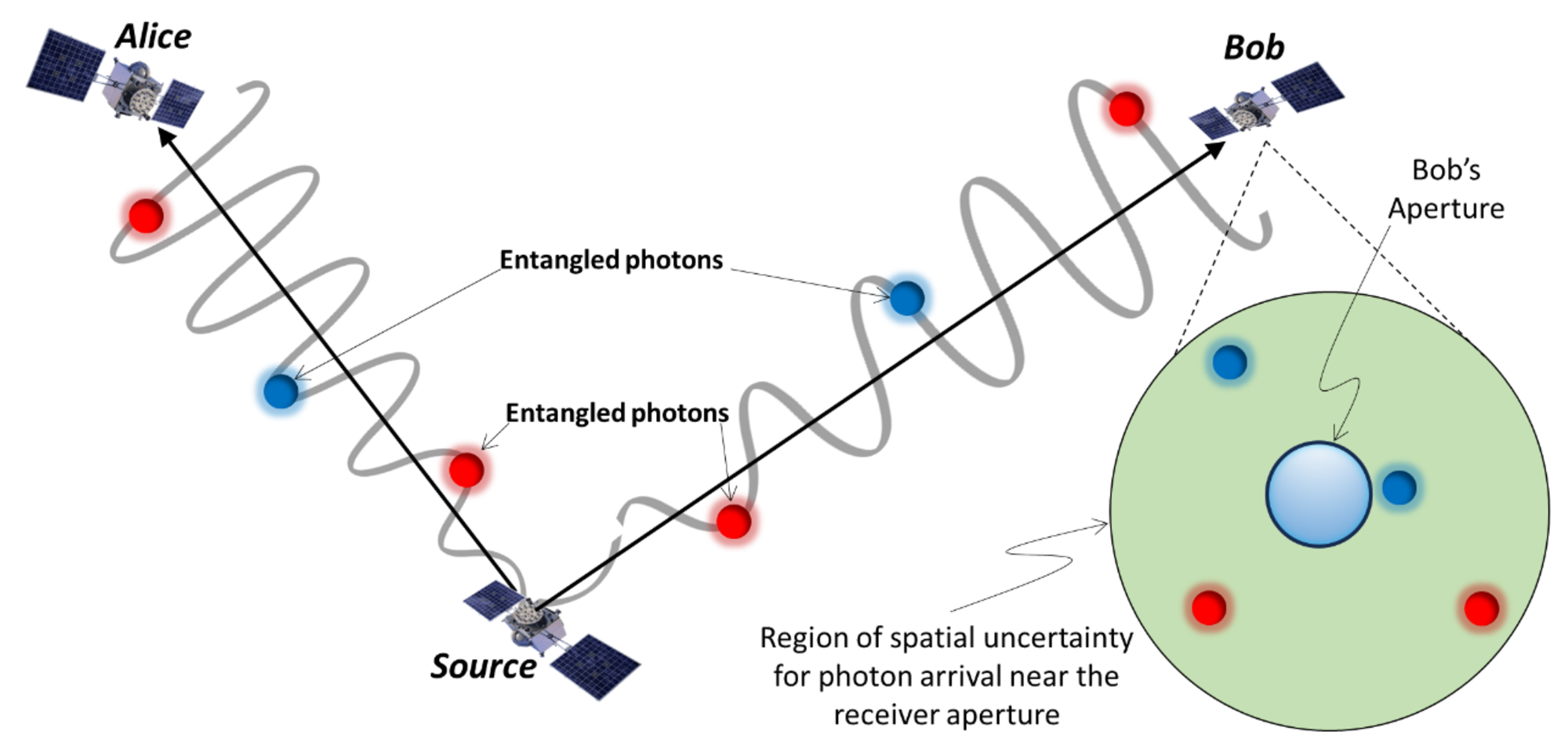}
		\caption{Quantum communication scheme based on the entanglement-based protocol for long-distance satellite links. Due to increasing link distances and corresponding channel parameters (e.g., beam divergence and pointing errors), the probability of entangled photon reception at the receiver’s aperture decreases significantly. A region of spatial uncertainty near the aperture illustrates the probabilistic nature of photon arrival.}
		\label{sm1}
	\end{center}
\end{figure}
%
\section{System Model} \label{sec:system_model}
In this section, we describe the photon-level model for inter-satellite entanglement-based QKD using the entanglement-based protocol. In this setting, a satellite node (referred to as the entangled photon source) generates polarization-entangled photon pairs and sends each member of the pair to two distant satellites, Alice and Bob. These receiving satellites act as QKD terminals performing polarization measurements on the received photons using randomly chosen measurement bases.

\subsection{Statistical Model of Entangled Photon Pair Generation}

The entangled photon source, positioned aboard a central relay satellite, produces polarization-entangled photon pairs using a nonlinear optical process such as spontaneous parametric down-conversion (SPDC). Ideally, each generated pair is in the maximally entangled Bell state:
\begin{align}
	|\Phi^+\rangle = \frac{1}{\sqrt{2}} \left( |H\rangle_A |H\rangle_B + |V\rangle_A |V\rangle_B \right),
\end{align}
where subscripts \( A \) and \( B \) denote the two distant receiver satellites, Alice and Bob, respectively.

In practice, the generation process is probabilistic and the number of entangled photon pairs produced within a given time slot of duration \( T_b \) is a random variable. Let \( n_t \) denote the number of photon pairs generated in a single time slot. For SPDC sources operating in the low-gain regime, this number can be accurately modeled as a Poisson-distributed random variable:
\begin{align} \label{df1}
	\mathbb{P}(n_t = k) = \frac{e^{-\mu_t} \mu_t^k}{k!}.
\end{align}
where \( \mu_t \) is the expected number of entangled photon pairs per time slot.

\subsection{Photon Modeling}
\label{subsec:photon_level_model}
Consider a single time slot of duration \( T_b \). For each channel \( q \in \{A,B\} \), let \( n_{r,q} \) denote the number of \emph{signal} photons (originally entangled photons destined for receiver \( q \)) that arrive at receiver \( q \). Let \( n_{b,q} \) be the number of \emph{background} photons in channel \( q \), arising from stray light or detector dark counts. The total received signal at satellite \( q \) can be expressed as the sum of all individual photons arriving at that detector:
\begin{align}
	S_{\mathrm{rec}}^{(q)} 
	\;=\; \sum_{i=1}^{n_{r,q}} \bigl|\psi_{s,r,i}^{(q)}\bigr\rangle
	\;+\;
	\sum_{j=1}^{n_{b,q}} \bigl|\psi_{b,j}^{(q)}\bigr\rangle.
\end{align}
where \( \bigl|\psi_{s,i}^{(q)}\bigr\rangle \) denotes the quantum state of the \( i \)-th received signal photon at receiver \( q \), and \( \bigl|\psi_{b,j}^{(q)}\bigr\rangle \) denotes the quantum state of the \( j \)-th background photon.
In the following, we model the statistical behavior of both the signal photons \( \bigl|\psi_{s,i}^{(q)}\bigr\rangle \) and the background photons \( \bigl|\psi_{b,j}^{(q)}\bigr\rangle \) in detail.

\subsubsection{Background Photons}
Each background photon \( \bigl|\psi_{b,j}^{(q)}\bigr\rangle \) is assumed to be unpolarized before reaching the receiver. For instance, in the rectilinear basis \(\{ |H\rangle, |V\rangle \}\), we model each background photon as
\begin{align}
	\bigl|\psi_{b,j}^{(q)}\bigr\rangle 
	\;=\; \frac{1}{\sqrt{2}}\, \bigl( |H\rangle + |V\rangle \bigr).
\end{align}
Upon passing through the receiver’s polarization analyzer (e.g., a polarizing beam splitter), the photon's polarization outcome collapses probabilistically into \(|H\rangle\) or \(|V\rangle\) with equal probability:
\begin{align}
	\bigl|\psi_{b,j}^{(q)}\bigr\rangle 
	~\rightarrow~
	\begin{cases}
		|H\rangle & \text{with probability } \tfrac{1}{2},\\[1em]
		|V\rangle & \text{with probability } \tfrac{1}{2}.
	\end{cases}
\end{align}
Hence, background photons introduce equal amounts of noise in all detection outcomes.

The background photon count \( n_{b,q} \) in channel \( q \) follows a Poisson distribution with mean \( \mu_{b,q} \), which depends on the detector dark counts, stray light intensity, and the receiver’s field-of-view (FoV). Specifically, for a time slot of duration \( T_b \),
\begin{align}
	n_{b,q} \sim \mathrm{Poisson}(\mu_{b,q}),
\end{align}
where
\begin{align}
	\mu_{b,q} = \Phi_{b,q} \, \Omega_{\mathrm{FoV},q} \, T_b,
\end{align}
with \(\Phi_{b,q}\) denoting the background photon flux density, and \(\Omega_{\mathrm{FoV},q}\) representing the receiver’s field-of-view solid angle, given by
\begin{align}
	\Omega_{\mathrm{FoV},q} = 2\pi \left(1 - \cos\left(\frac{\theta_{\mathrm{FoV},q}}{2}\right)\right),
\end{align}
where \(\theta_{\mathrm{FoV},q}\) is the receiver’s field-of-view angle. 
Since the background photons are assumed unpolarized, they contribute uniformly to all polarization measurements in each channel \( q \).

\subsubsection{Signal Photon}
The signal photons \( \bigl|\psi_{s,i}^{(q)}\bigr\rangle \), corresponding to the entangled photons generated by the source, are transmitted over a free-space optical link to receiver \( q \in \{A, B\} \).
We assume that each channel \( q \) is characterized by an effective transmittance factor \( \eta_q \), which accounts for all channel impairments such as geometric loss, pointing error, and receiver detection efficiency. The detailed modeling of \( \eta_q \) will be presented in the next section.


\section{Channel Transmittance Model}
\label{sec:channel_model}
In this section, we present the channel transmittance model for the inter-satellite free-space optical links used in the entanglement-based QKD system. The system consists of two independent free-space optical (FSO) links connecting the entangled photon source to the two receiver satellites, Alice and Bob. We denote these links by \(q \in \{A, B\}\), corresponding to the channels associated with Alice and Bob, respectively. 


\subsection{Photon Propagation}
Photons are transmitted along the \(z_q\)-axis from the entangled photon source to the receiver of channel \(q \in \{A, B\}\), located at a distance \(Z_{L,q}\). Due to imperfect tracking systems, the emitted Gaussian beam may deviate from its ideal propagation axis \cite{dabiri2019tractable,dabiri2025novel}. The transmitter angular tracking error for link \(q\) is denoted by the random variable \( \theta_{t,q} = (\theta_{tx,q}, \theta_{ty,q}) \), where \( \theta_{tx,q}, \theta_{ty,q} \sim \mathcal{N}(0, \sigma_{\theta,t,q}^2) \).
At the receiver plane \(z = Z_{L,q}\), the beam center is displaced in the transverse \((x,y)\) plane by:
\begin{align}
	(x_{0,q}, y_{0,q}) = (Z_{L,q} \theta_{tx,q}, Z_{L,q} \theta_{ty,q}).
\end{align}


The beam intensity profile at the receiver is modeled as a normalized Gaussian distribution with beam width (standard deviation) \(w_{z,q}\), which depends on the transmission distance \(Z_{L,q}\), the operating wavelength \(\lambda\), and the initial beam waist \(w_{0,q}\). Specifically, the beam width at the receiver plane for channel \(q\) is given by:
\begin{align}
	w_{z,q} = w_{0,q} \sqrt{1 + \left( \frac{Z_{L,q}}{z_{R,q}} \right)^2 },
\end{align}
where \(z_{R,q}\) denotes the Rayleigh range for link \(q\) and is expressed as:
\begin{align}
	z_{R,q} = \frac{\pi w_{0,q}^2}{\lambda}.
\end{align}
In typical inter-satellite communication scenarios, where the propagation distance is much larger than the Rayleigh range (i.e., \(Z_{L,q} \gg z_{R,q}\)), the beam width can be approximated as:
\begin{align}
	w_{z,q} \approx \frac{\lambda Z_{L,q}}{\pi w_{0,q}}.
\end{align}
This expression highlights the direct impact of the transmission distance \(Z_{L,q}\) and the wavelength \(\lambda\) on the beam divergence at the receiver. The conditional probability density function (pdf) of a photon's transverse position \((x, y)\), given the pointing error \(\theta_{t,q}\), is:
\begin{align}
	p(x, y \mid \theta_{t,q}) = \frac{1}{2\pi w_{z,q}^2} \exp\left( -\frac{(x-x_{0,q})^2 + (y-y_{0,q})^2}{2w_{z,q}^2} \right).
\end{align}

\subsection{Aperture Coupling Efficiency}
Assume that the receiver of channel \(q\) has a circular aperture of radius \(a_q\). The conditional probability that a photon is successfully collected by the receiver, given the pointing error \(\theta_{t,q}\), is:
\begin{align}
	\eta'_p(\theta_{t,q}) = \iint_{x^2+y^2 \leq a_q^2} p(x, y \mid \theta_{t,q}) \, dx \, dy.
\end{align}
For typical satellite communication scenarios where \(w_{z,q} \gg a_q\), the above integral can be approximated as \cite{dabiri2024modulating1,dabiri2024modulating}:
\begin{align} \label{sd1_revised}
	\eta'_p(\theta_{t,q}) \approx \frac{a_q^2}{2w_{z,q}^2} \exp\left( -\frac{Z_{L,q}^2(\theta_{tx,q}^2+\theta_{ty,q}^2)}{2w_{z,q}^2} \right).
\end{align}

Since \(\theta_{tx,q}\) and \(\theta_{ty,q}\) are independent Gaussian random variables, the squared radial displacement \(r_q^2 = \theta_{tx,q}^2 + \theta_{ty,q}^2\) follows an exponential distribution with pdf:
\begin{align}
	f_{r_q^2}(r_q^2) = \frac{1}{2\sigma_{\theta,t,q}^2} \exp\left( -\frac{r_q^2}{2\sigma_{\theta,t,q}^2} \right), \quad r_q^2 \geq 0.
\end{align}
Substituting \(r_q^2\) into the expression for \(\eta'_p\), we obtain:
\begin{align}
	\eta'_p = C_q \exp(-\lambda_q r_q^2),
\end{align}
where \(C_q = \frac{a_q^2}{2w_{z,q}^2}\) and \(\lambda_q = \frac{Z_{L,q}^2}{2w_{z,q}^2}\).
The probability density function of \(\eta'_p\) is therefore given by:
\begin{align} \label{sd3_revised}
	f_{\eta'_p}(\eta'_p) = \gamma_q \left( \frac{2w_{z,q}^2}{a_q^2} \right)^{\gamma_q} {\eta'_p}^{\gamma_q-1}, \quad 0 \leq \eta'_p \leq \frac{a_q^2}{2w_{z,q}^2},
\end{align}
where \(\gamma_q = \frac{w_{z,q}^2}{4Z_{L,q}^2 \sigma_{\theta,t,q}^2}\). 

However, not all photons collected by the receiver aperture are ultimately registered by the quantum detector. Due to practical hardware constraints and to mitigate background noise, the FoV of the optical receiver is typically limited by a narrow spatial filter or optical concentrator. 
To reduce the background photon rate, the FoV angle \( \theta_{\mathrm{FoV},q} \) must be carefully limited. A smaller FoV leads to a smaller solid angle \( \Omega_{\mathrm{FoV},q} \), which in turn reduces the mean number of background photons \( \mu_{b,q} \). However, this strict limitation on the FoV imposes a trade-off: signal photons that deviate from the optical axis due to angular jitter may also fall outside the detection cone and get rejected, even if they pass through the aperture.

This effect becomes especially significant in the presence of angular misalignment at the receiver. Let \( \theta_{\mathrm{FoV},q} \) denote the FoV angle, and let \( \sigma_{\mathrm{FoV},q} \) represent the standard deviation of the receiver’s instantaneous angular deviation. The receiver angular error is assumed to follow a Rayleigh distribution, modeling the magnitude of two-dimensional angular jitter \cite{dabiri2018channel}:
\begin{align}
	f_{\Theta_q}(\theta) = \frac{\theta}{\sigma_{\mathrm{FoV},q}^2} \exp\left( -\frac{\theta^2}{2\sigma_{\mathrm{FoV},q}^2} \right), \quad \theta \geq 0.
\end{align}
The probability that an incoming photon lies within the receiver's FoV cone is equivalent to the probability that the angular displacement \( \theta \) is less than the FoV limit \( \theta_{\mathrm{FoV},q} \). This cumulative probability is given by:
\begin{align}
	P_{\mathrm{FoV},q} = \mathbb{P}(\theta \leq \theta_{\mathrm{FoV},q}) = 1 - \exp\left( -\frac{ \theta_{\mathrm{FoV},q}^2 }{ 2 \sigma_{\mathrm{FoV},q}^2 } \right).
\end{align}
Therefore, the overall probability that a photon successfully reaches the quantum detector, after passing through the aperture and satisfying the FoV constraint, is:
\begin{align} \label{kh1}
	\eta_{p,q} = \eta'_p \cdot P_{\mathrm{FoV},q}.
\end{align}
Using \eqref{sd3_revised} and \eqref{kh1}, the resulting PDF of \( \eta_{p,q} \) also follows a scaled power-law:
\begin{align}
	f_{\eta_{p,q}}(\eta_{p,q}) &= \gamma_q \left( \frac{2w_{z,q}^2}{a_q^2 P_{\mathrm{FoV},q}^2} \right)^{\gamma_q} \eta_{p,q}^{\gamma_q - 1}, \nonumber \\
	&~~~~~~~ 0 \leq \eta_{p,q} \leq \frac{a_q^2}{2w_{z,q}^2} P_{\mathrm{FoV},q},
\end{align}
This distribution will be used in the following sections for modeling signal detection and key rate analysis.

\section{Photon Reception and Secret Key Rate Analysis}
\label{sec:key_rate_analysis}
In this section, we analyze the photon reception statistics and the resulting secret key rate for the entanglement-based QKD system described in Section~\ref{sec:channel_model}. 

\subsection{Photon Reception Rate per Channel}
Recall from Section~\ref{sec:channel_model} that the aperture coupling efficiency of channel \(q\) is denoted by \(\eta_{p,q}\), which accounts for the geometric loss and pointing error effects. Additionally, let \(\eta_{\mathrm{det},q}\) represent the detection efficiency of the receiver on channel \(q\). Therefore, the average number of received signal photons at receiver \(q\) per time slot is given by:
\begin{align} \label{sd2}
	\eta_{r,q} = \mu_t \eta_q = \mu_t \eta_{p,q} \eta_{\mathrm{det},q},
\end{align}
where \(\mu_t\) is the mean photon pair generation rate of the entangled source.
Based on \eqref{sd1_revised} and \eqref{sd2}, \(\eta_{r,q}\) is a function of the angular tracking error variance \(\sigma_{\theta,t,q}^2\) and the link distance \(Z_{L,q}\). In the typical inter-satellite regime, where both the link distance and pointing errors are significant, \(\eta_{r,q}\) becomes much smaller than unity, i.e., \(\eta_{r,q} \ll 1\).

Let \(c_q = \mu_t \eta_{\mathrm{det},q}\), so that \(\eta_{r,q} = c_q \eta_{p,q}\).
Using the standard change-of-variables formula, the PDF of \(\eta_{r,q}\) is given by:
\begin{align} \label{sb7}
	f_{\eta_{r,q}}(\eta_{r,q}) = \frac{1}{c_q} f_{\eta_{p,q}}\left( \frac{\eta_{r,q}}{c_q} \right).
\end{align}
Substituting from \eqref{sd3_revised}, we obtain:
\begin{align} \label{sd4}
	f_{\eta_{r,q}}(\eta_{r,q}) 
	= \gamma_q \left( \frac{2w_{z,q}^2}{a_q^2 c_q^2} \right)^{\gamma_q/2} \eta_{r,q}^{\gamma_q-1},
\end{align}
for \( 0 \leq \eta_{r,q} \leq \frac{a_q^2}{2w_{z,q}^2} \times c_q \).

Given the number of generated entangled photon pairs \( n_t \) within a time slot of duration \( T_b \), the number of received signal photons \( n_{r,q} \) at receiver \( q \in \{A, B\} \) follows a Binomial distribution, since each transmitted photon independently survives the channel with probability \( \eta_q \) (including geometric loss, pointing error, and detection efficiency) \cite{papoulis2002probability}. Specifically,
\begin{align}
	n_{r,q} \mid n_t \sim \mathrm{Binomial}(n_t, \eta_q),
\end{align}
with the probability mass function (PMF) given by
\begin{align}
	\mathbb{P}(n_{r,q}=k \mid n_t) = \binom{n_t}{k} \eta_q^k (1-\eta_q)^{n_t-k},
\end{align}
where \( k=0,1,\dots,n_t \) and \( \eta_q \in [0,1] \).

In entanglement-based QKD systems, the average photon pair generation rate \( \mu_t \) is typically very small (\( \mu_t \ll 1 \)) to minimize multi-pair events. Under this low-brightness regime, and considering that \( n_t \) is Poisson-distributed with mean \(\mu_t\), the resulting distribution of \(n_{r,q}\) can be accurately approximated by a Poisson distribution with mean \(\eta_{r,q}\), i.e.,
\begin{align}
	n_{r,q} \sim \mathrm{Poisson}(\eta_{r,q}),
\end{align}
where \(\eta_{r,q}\) is defined in \eqref{sd2}. This approximation significantly simplifies the subsequent performance analysis while maintaining high accuracy in typical QKD scenarios.

\subsection{Secret Key Generation Rate}
In the entanglement-based protocols like BBM92, a raw key bit is successfully generated if the following conditions are simultaneously satisfied:
\begin{itemize}
	\item A photon is successfully detected at both receivers \(A\) and \(B\) within the same time slot.
	\item Both receivers select the measurement basis designated for key generation.
\end{itemize}
Note that both signal photons and background photons are considered. Specifically, given a realization of the channel parameters, the number of detected signal photons \(n_{r,q}\) at receiver \(q\) per time slot follows a Poisson distribution with mean \(\eta_{r,q}\), where \(\eta_{r,q}\) is defined in \eqref{sd2}. Similarly, the number of background photons \(n_{b,q}\) detected at receiver \(q\) follows an independent Poisson distribution with mean \(\mu_{b,q}\).
Thus, the total number of detected photons at receiver \(q\) is
\begin{align}
	n_{\mathrm{tot},q} = n_{r,q} + n_{b,q}.
\end{align}
Since the sum of two independent Poisson random variables is also Poisson-distributed with the sum of their means \cite{papoulis2002probability}, we have
\begin{align}
	n_{\mathrm{tot},q} \sim \mathrm{Poisson}(\eta_{r,q} + \mu_{b,q}).
\end{align}

Given realizations of \(\eta_{r,A}\) and \(\eta_{r,B}\), the instantaneous secret key generation rate per time slot can be approximated as
\begin{align}
	R_{\mathrm{key}}(\eta_{r,A}, \eta_{r,B}) \simeq P_{\mathrm{basis}} \times P_{\mathrm{click},A} \times P_{\mathrm{click},B},
\end{align}
where \(P_{\mathrm{basis}}\) denotes the probability that Alice and Bob select the same measurement basis, and \(P_{\mathrm{click},q}\) represents the probability that receiver \(q\) registers at least one photon (either signal or background) within the time slot.
According to basic principles of probability theory \cite{papoulis2002probability}, the basis selection process in the BBM92 protocol can be modeled as two independent and uniform random choices, similar to tossing two fair coins. Consequently, the probability that Alice and Bob select the same basis is \(P_{\mathrm{basis}} = 1/2\).

Moreover, according to the properties of the Poisson distribution, the probability of detecting at least one photon at receiver \(q\) is
\begin{align}
	P_{\mathrm{click},q} = 1 - \exp(-\eta_{r,q} - \mu_{b,q}).
\end{align}
In the practical regime where \(\eta_{r,q}, \mu_{b,q} \ll 1\), which is typical for entanglement-based QKD systems, we can approximate
\begin{align}
	P_{\mathrm{click},q} \simeq \eta_{r,q} + \mu_{b,q},
\end{align}
using the first-order Taylor expansion of the exponential function.
Therefore, the secret key generation rate can be approximated by
\begin{align}
	R_{\mathrm{key}}(\eta_{r,A}, \eta_{r,B}) \simeq \frac{1}{2} (\eta_{r,A} + \mu_{b,A}) (\eta_{r,B} + \mu_{b,B}).
\end{align}

Since \(\eta_{r,A}\) and \(\eta_{r,B}\) are random variables depending on the channel pointing errors, and \(\mu_{b,A}\) and \(\mu_{b,B}\) are fixed background noise rates, the average secret key generation rate is given by
\begin{align}
	\bar{R}_{\mathrm{key}} = \frac{1}{2} \left( \mathbb{E}\left[\eta_{r,A}\right] + \mu_{b,A} \right) \left( \mathbb{E}\left[\eta_{r,B}\right] + \mu_{b,B} \right),
\end{align}
where the expected value \(\mathbb{E}[\eta_{r,q}]\) can be computed as
\begin{align}
	\mathbb{E}[\eta_{r,q}] = \int_{0}^{c_q \frac{a_q^2}{2w_{z,q}^2}} \eta_{r,q} f_{\eta_{r,q}}(\eta_{r,q}) \, d\eta_{r,q},
\end{align}
where \(c_q = \mu_t \eta_{\mathrm{det},q}\) and \(f_{\eta_{r,q}}(\eta_{r,q})\) is the scaled power-law PDF derived in \eqref{sd4}.
Substituting the expression of \(f_{\eta_{r,q}}(\eta_{r,q})\) from \eqref{sd4} and evaluating the integral yields
\begin{align}
	\mathbb{E}[\eta_{r,q}] = \mu_t \eta_{\mathrm{det},q} \times \frac{\gamma_q a_q^2}{2(\gamma_q+1) w_{z,q}^2},
\end{align}
where \(\gamma_q = \frac{w_{z,q}^2}{4Z_{L,q}^2\sigma_{\theta,t,q}^2}\).

\subsection{Error Probability and QBER Analysis}
In a typical entanglement-based QKD system, errors in the generated key bits arise from background photon detections and multi-pair emissions. A raw key bit is generated only when both receivers detect photons and select the same measurement basis. Therefore, the relevant error probabilities must account for both photon detection events and basis matching probability.
When both detected photons originate from the same entangled photon pair, the measurement outcomes are either perfectly correlated or anti-correlated depending on the Bell state, resulting in an almost negligible error probability. This is because, in space-based quantum communication, photons propagate through a vacuum without significant depolarization or channel-induced distortions. Therefore, the contribution of true entangled photon detections to the overall error is negligible compared to errors caused by background photons and multi-pair emissions.

The main error sources are classified into three categories:
\begin{itemize}
	\item \textbf{(E1)} Both clicks originate from background photons.
	\item \textbf{(E2)} One click from signal photon, one click from background photon.
	\item \textbf{(E3)} Multi-pair emission causes mismatched clicks.
\end{itemize}
Here, a click refers to a successful photon detection event registered by the quantum receiver. 
The corresponding error probabilities and event occurrence probabilities are as follows.

\subsubsection{E1 (Both clicks from background photons)}
In this case, both receivers detect photons that originate from uncorrelated background noise sources. Since background photons are unpolarized and uncorrelated, the measurement outcomes are random, resulting in an error probability of \( \frac{1}{2} \). The occurrence probability of this event, including the basis matching factor, is given by
\begin{align} \label{sb1}
	P_{E1} = \frac{1}{2} \mu_{b,A} \mu_{b,B}.
\end{align}

\subsubsection{E2 (One signal photon and one background photon)}
This event occurs when one receiver detects a signal photon from the entangled source while the other receiver detects a background photon. Since the background photon provides no useful correlation, the probability of measurement mismatch remains \( \frac{1}{2} \). The occurrence probability of this event, accounting for both possible receiver combinations and basis matching, is
\begin{align} \label{sb21}
	P_{E2} = \frac{1}{2} (\eta_{r,A} \mu_{b,B} + \mu_{b,A} \eta_{r,B}).
\end{align}

\subsubsection{E3 (Multi-pair emission events)}
Based on \eqref{df1}, the number of generated entangled photon pairs per time slot, denoted by \(n_t\), follows a Poisson distribution with mean \(\mu_t\). In the following, we analyze the cases where two and three photon pairs are generated, as they represent the dominant contributions to multi-pair events. Higher-order pair generation events occur with significantly lower probabilities and can thus be safely neglected without appreciable loss of accuracy.

The probability of generating exactly two photon pairs within a time slot is given by
\begin{align}
	\mathbb{P}(n_t = 2) = \frac{e^{-\mu_t} \mu_t^2}{2}.
\end{align}
In the event of two photon pairs being generated, there are four photons in total: two directed toward Alice and two toward Bob. An error occurs if the detected photons at Alice and Bob originate from different entangled pairs, leading to uncorrelated measurement outcomes. However, for an error to occur, the photons must first survive transmission and detection, which happen independently with probabilities \(\eta_{r,A}\) and \(\eta_{r,B}\), respectively.
Considering the possible detection combinations, an error occurs if the photons detected at Alice and Bob belong to different entangled pairs, which occurs with probability \( \frac{1}{2} \). Furthermore, assuming independent detection events, the conditional probability of an error given a two-pair event can be approximated as proportional to \(\eta_{r,A} \eta_{r,B}\). Therefore, the effective error contribution from two-pair events becomes
\begin{align} \label{sf1}
	P_{\text{error, 2-pair}} = \frac{1}{2} \times \frac{1}{2} \times \frac{e^{-\mu_t} \mu_t^2}{2} \times \eta_{r,A} \eta_{r,B},
\end{align}
where the first factor \( \frac{1}{2} \) accounts for pair mismatch and the second \( \frac{1}{2} \) represents the measurement basis error. 

Similarly, the probability of generating exactly three photon pairs is
\begin{align}
	\mathbb{P}(n_t = 3) = \frac{e^{-\mu_t} \mu_t^3}{6}.
\end{align}
Here, six photons are generated (three toward each receiver). If Alice and Bob detect photons originating from different entangled pairs, which occurs with probability \( \frac{2}{3} \), an error can occur. Accounting again for independent detection and basis mismatch, the effective error contribution from three-pair events is
\begin{align}  \label{sf2}
	P_{\text{error,3-pair}} = \frac{2}{3} \times \frac{1}{2} \times \frac{e^{-\mu_t} \mu_t^3}{6} \times \eta_{r,A} \eta_{r,B}.
\end{align}

The total contribution of multi-pair emissions to the secret key error probability can be obtained by summing the individual contributions from two-pair and three-pair generation events. Substituting from \eqref{sf1} and \eqref{sf2}, we have
\begin{align} \label{sb3}
	P_{\text{multi-pair}} 
	= \frac{e^{-\mu_t} \eta_{r,A} \eta_{r,B}}{72} \left(9 \mu_t^2 + 4 \mu_t^3\right).
\end{align}
Given that higher-order multi-pair events (i.e., more than three pairs) occur with significantly lower probabilities when \(\mu_t\) is small, their contributions can be neglected without substantial loss of accuracy.
Table~\ref{tab:error_events} summarizes the key error events that affect the secret key generation process, along with their associated error probabilities (EP) and occurrence probabilities.

\begin{table*}[!t]
	\centering
	\caption{Summary of Error Events (EP: Error Probability)}
	\label{tab:error_events}
	\begin{tabular}{|c|l|c|c|}
		\hline
		\textbf{Event Type} & \textbf{Description} & \textbf{EP} & \textbf{Occurrence Probability} \\ \hline
		Both clicks from background & Both Alice and Bob detect background photons & \( 1/2 \) & \( \frac{1}{2} \mu_{b,A} \mu_{b,B} \) \\ \hline
		One signal, one background & One receiver detects a signal, the other detects background & \( 1/2 \) & \( \frac{1}{2} (\eta_{r,A} \mu_{b,B} + \mu_{b,A} \eta_{r,B}) \) \\ \hline
		Two-pair multi-photon events & Error from two entangled pairs generated & \( 1/2 \) & \( \frac{1}{8} e^{-\mu_t} \mu_t^2 \eta_{r,A} \eta_{r,B} \) \\ \hline
		Three-pair multi-photon events & Error from three entangled pairs generated & \( 1/2 \) & \( \frac{1}{18} e^{-\mu_t} \mu_t^3 \eta_{r,A} \eta_{r,B} \) \\ \hline
		Higher-order events (four or more pairs) & Negligible compared to other cases; ignored & -- & Negligible \\ \hline
	\end{tabular}
\end{table*}

Consequently, the conditional expected number of erroneous bits per time slot, given the instantaneous values of \(\eta_{r,A}\) and \(\eta_{r,B}\), denoted by \( P_{\mathrm{err}}(\eta_{r,A}, \eta_{r,B}) \), can be written as \eqref{sb2}. 
\begin{figure*}[!t]
	\begin{align} \label{sb2}
		P_{\mathrm{err}}(\eta_{r,A}, \eta_{r,B}) = \frac{1}{2} \mu_{b,A} \mu_{b,B} + \frac{1}{2} (\eta_{r,A} \mu_{b,B} + \mu_{b,A} \eta_{r,B}) + 
		\frac{1}{2}\eta_{r,A} \eta_{r,B} 
		 \left( \frac{1}{8} e^{-\mu_t} \mu_t^2 + \frac{1}{18} e^{-\mu_t} \mu_t^3 \right).
	\end{align}
	\hrulefill
	\vspace*{0pt}
\end{figure*}
Similarly, the conditional expected number of sifted key bits per time slot, given the instantaneous values of \(\eta_{r,A}\) and \(\eta_{r,B}\), is given by
\begin{align}
	P_{\mathrm{click}}(\eta_{r,A}, \eta_{r,B}) = \frac{1}{2} (\eta_{r,A} + \mu_{b,A})(\eta_{r,B} + \mu_{b,B}).
\end{align}
Thus, the conditional QBER, given the instantaneous values of \(\eta_{r,A}\) and \(\eta_{r,B}\), can be expressed as
\begin{align}
	\text{QBER}(\eta_{r,A}, \eta_{r,B}) = \frac{   P_{\mathrm{err}}(\eta_{r,A}, \eta_{r,B})    }
	{P_{\mathrm{click}}(\eta_{r,A}, \eta_{r,B})},
\end{align}
which simplifies to \eqref{sb6}.
\begin{figure*}[!t]
	\begin{align} \label{sb6}
		\text{QBER}(\eta_{r,A}, \eta_{r,B}) = \frac{ \mu_{b,A} \mu_{b,B} +  \eta_{r,A} \mu_{b,B} + \mu_{b,A} \eta_{r,B} 
			+ \frac{1}{8} e^{-\mu_t} \mu_t^2 \eta_{r,A} \eta_{r,B} 
			+ \frac{1}{18} e^{-\mu_t} \mu_t^3 \eta_{r,A} \eta_{r,B} }{  (\eta_{r,A} + \mu_{b,A})(\eta_{r,B} + \mu_{b,B}) }.
	\end{align}
	\hrulefill
	\vspace*{0pt}
\end{figure*}

Finally, the average QBER can be obtained by integrating over the joint distribution of the random photon reception rates \(\eta_{r,A}\) and \(\eta_{r,B}\), as
\begin{align}
	& \overline{\text{QBER}} = \gamma_A \gamma_B \left( \frac{2w_{z,A}^2}{a_A^2 c_A^2} \right)^{\gamma_A} 
	\times
	\left( \frac{2w_{z,B}^2}{a_B^2 c_B^2} \right)^{\gamma_B}  \times \nonumber \\
	&  \iint\limits_{\eta_{r,A},\eta_{r,B}} \text{QBER}(\eta_{r,A}, \eta_{r,B}) \cdot 
	 \eta_{r,A}^{\gamma_A - 1}   \eta_{r,B}^{\gamma_B - 1}
	\, d\eta_{r,A} \, d\eta_{r,B},
\end{align}
where \(c_q = \mu_t \eta_{\mathrm{det},q}\), and \(\gamma_q = \frac{w_{z,q}^2}{4 Z_{L,q}^2 \sigma_{\theta,t,q}^2}\), for \(q \in \{A, B\}\).

\begin{table}[!t]
	\centering
	\caption{Default simulation parameters used for both channels \(q \in \{A, B\}\).}
	\label{tab:values}
	\begin{tabular}{|c|c|}
		\hline
		\textbf{Symbol} & \textbf{Value} \\ \hline
		\( \lambda \) & \( 1550 \, \mathrm{nm} \) \\ \hline
		\( w_0 \) & \( 0.08 \, \mathrm{m} \) \\ \hline
		\( \mu_t \) & \( 0.05 \) \\ \hline
		\( T_{\mathrm{bit}} \) & \( 0.1 \, \mathrm{ns} \) \\ \hline
		\( \eta_{\mathrm{det},q} \) & \( 0.6 \) \\ \hline
		\( \Phi_{b,q} \) & \( 10^7 \, \mathrm{photons/s/sr} \) \\ \hline
		\( \theta_{\mathrm{FoV},q} \) & \( 1 \, \mathrm{mrad} \) \\ \hline
		\( \sigma_{\mathrm{FoV},q} \) & \( 100 \, \mu\mathrm{rad} \) \\ \hline
		\( a_q \) & \( 0.15 \, \mathrm{m} \) \\ \hline
		\( Z_{L,q} \) & \( 100 \text{ km} \) to \( 1000 \text{ km} \) \\ \hline
		\( \sigma_{\theta,t,q}^2 \) & \( (3 \, \mu\mathrm{rad})^2 \) \\ \hline
	\end{tabular}
\end{table}

\section{Simulation Results}
In this section, we evaluate the performance of the entanglement-based quantum communication system over long-distance satellite links. Specifically, we investigate the impact of various physical and system-level parameters on the quantum key generation rate and the QBER. These parameters include the background photon flux density \( \Phi_{b,q} \), transmitter angular tracking precision (via \( \sigma_{\theta,t,q}^2 \)), receiver FoV angle \( \theta_{\mathrm{FoV},q} \), angular jitter variance \( \sigma_{\mathrm{FoV},q}^2 \), link distance \( Z_{L,q} \), and the mean pair generation rate \( \mu_t \).
The default values used for the simulations are listed in Table~\ref{tab:values}, which are chosen to reflect realistic configurations commonly found in satellite-based quantum optical systems. Unless otherwise stated, all figures assume these default values; any deviations or parameter sweeps are explicitly noted in the respective figure captions.

\begin{figure*}
	\centering
	\subfloat[] {\includegraphics[width=3.45 in]{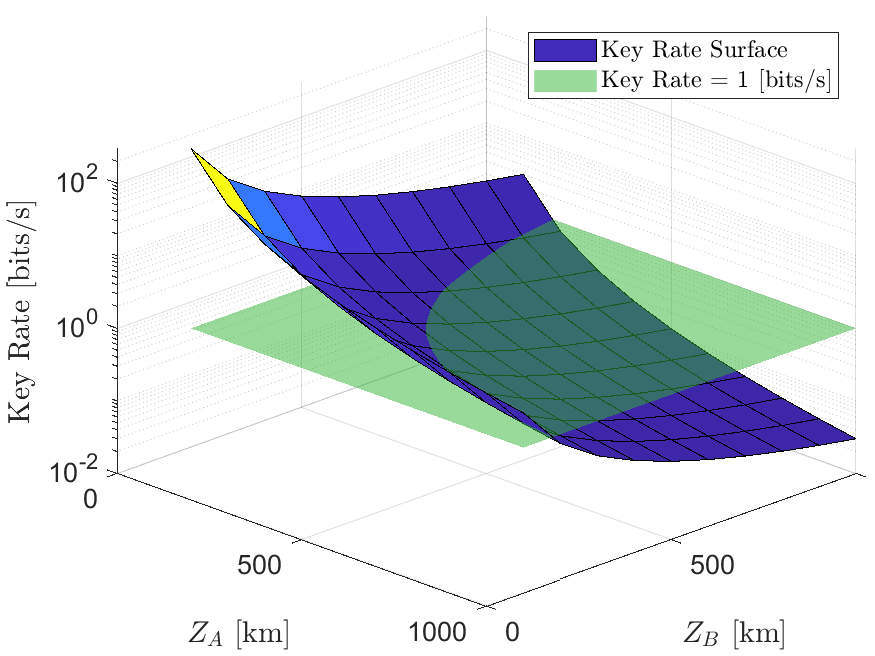}
		\label{ch1}
	}
	\hfill
	\subfloat[] {\includegraphics[width=3.45 in]{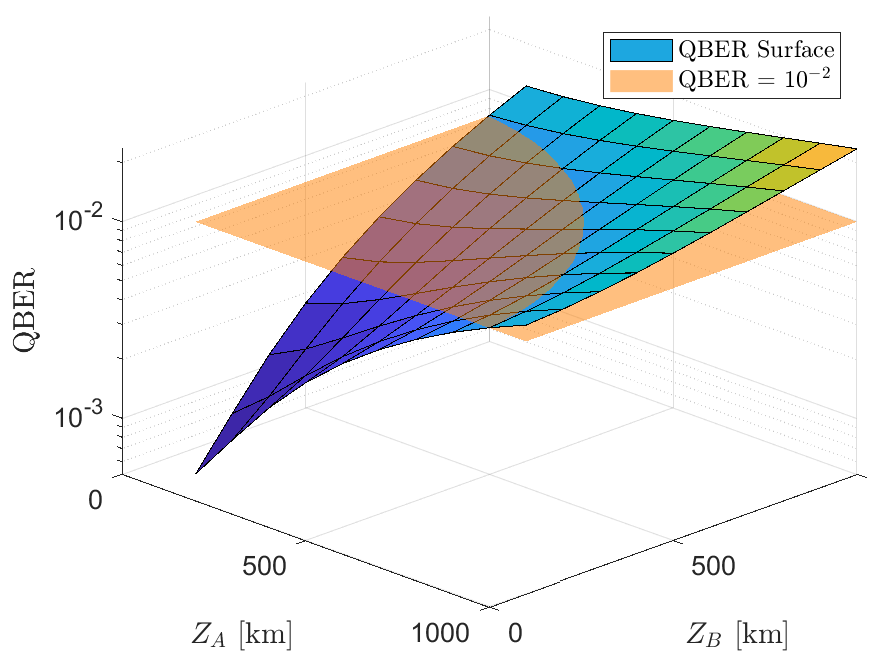}
		\label{ch2}
	}
\caption{
	Performance evaluation of the entanglement-based QKD system over long-distance satellite links. (a) Average secret key rate as a function of the dynamic link distances \(Z_A\) and \(Z_B\); (b) Corresponding QBER. Due to rapid changes in satellite positions, the inter-satellite link lengths vary significantly during each pass. The results indicate that at longer distances, both the key rate drops below 1 bit/s and the QBER exceeds acceptable limits (e.g., \(10^{-2}\)), highlighting a critical trade-off in maintaining reliable quantum communication over dynamically changing orbits.
}

	\label{ch}
\end{figure*}
%

\begin{figure}
	\centering
	\subfloat[] {\includegraphics[width=3.35 in]{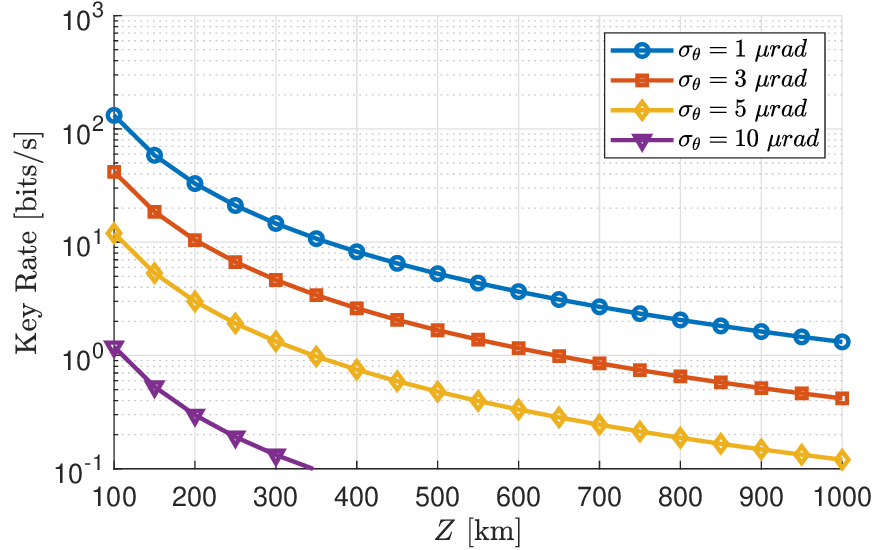}
		\label{cb1}
	}
	\hfill
	\subfloat[] {\includegraphics[width=3.35 in]{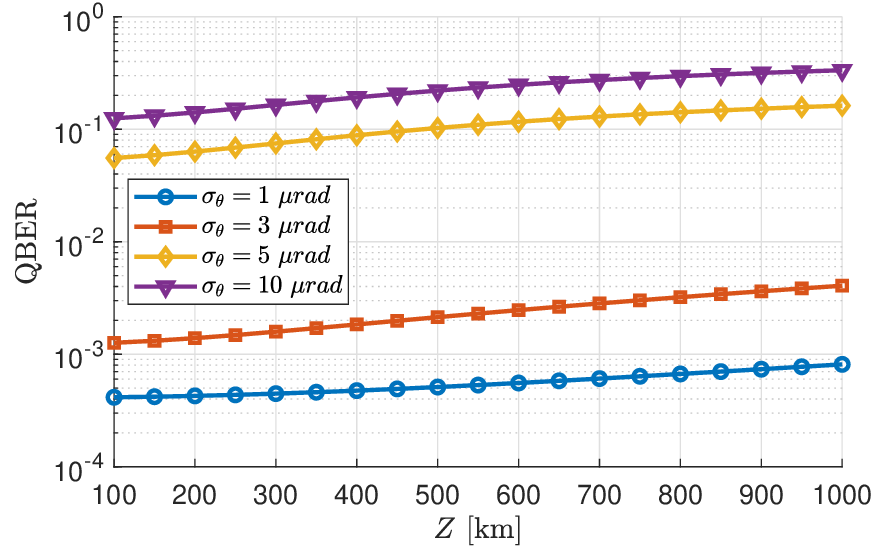}
		\label{cb2}
	}
	\caption{
		Impact of transmitter angular tracking accuracy on QKD performance. (a) Average secret key rate versus Alice's link distance, for various values of transmitter angular jitter \( \sigma_{\theta,t} \in \{1, 3, 5, 10\} \, \mu\mathrm{rad} \), with Bob's link fixed at \(500\,\mathrm{km}\). (b) Corresponding QBER performance.
	}
	
	\label{cb}
\end{figure}
%

\begin{figure}
	\centering
	\subfloat[] {\includegraphics[width=3.35 in]{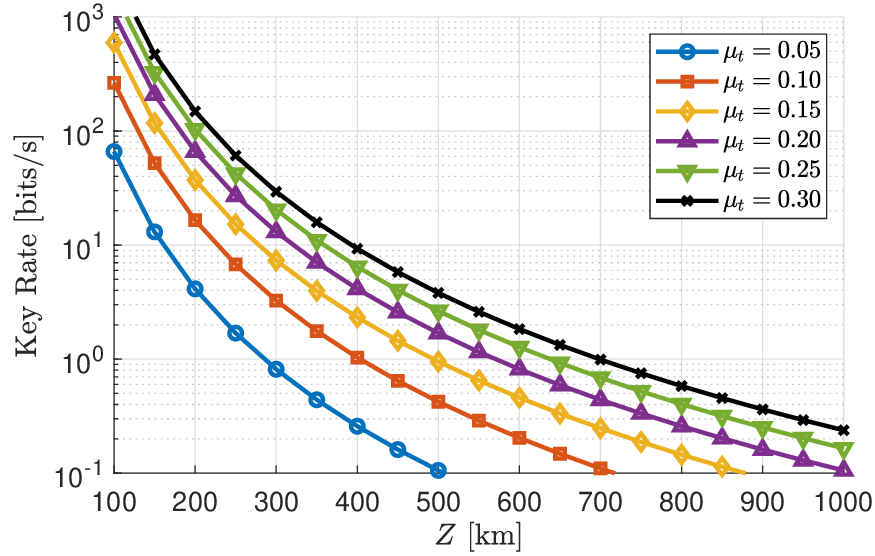}
		\label{cn1}
	}
	\hfill
	\subfloat[] {\includegraphics[width=3.35 in]{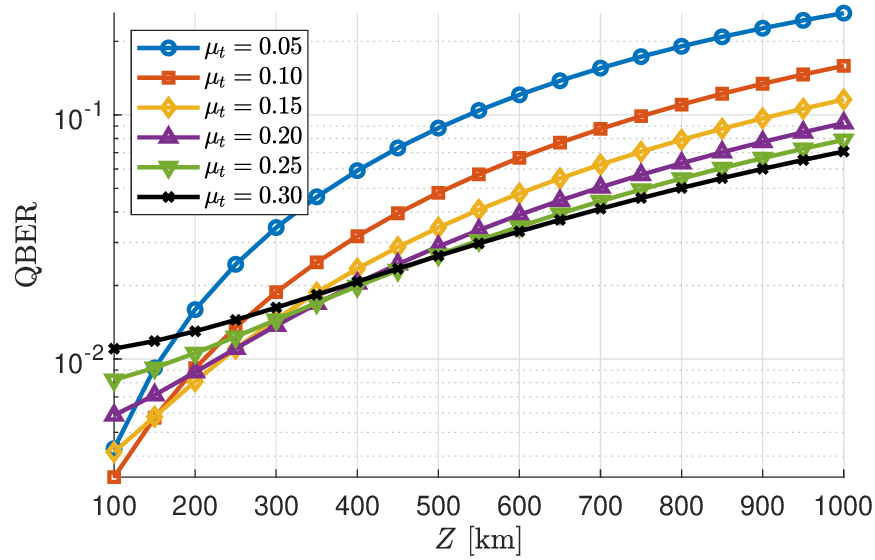}
		\label{cn2}
	}
	\caption{
		Impact of photon pair generation rate \(\mu_t\) on system performance. (a) Average secret key rate versus Alice's link distance for various \(\mu_t\) values in the range \(0.05\) to \(0.3\), with Bob's link fixed at \(500\,\mathrm{km}\). (b) Corresponding QBER values. While increasing \(\mu_t\) improves the probability of successful photon reception, it also leads to higher QBER due to multi-pair emissions. This highlights the critical role of \(\mu_t\) selection in system design.	}
	
	\label{cn}
\end{figure}
%

The most prominent characteristic of satellite-based quantum communication is the long and dynamically varying link distances. Due to the high orbital velocities of satellites and the continuous switching of communication paths, the optical link distances between nodes (e.g., Alice and Bob) vary rapidly and non-uniformly during each pass. 
To analyze the joint effect of this variability, Fig.~\ref{ch1} and Fig.~\ref{ch2} illustrate the average secret key rate and the average QBER, respectively, as functions of the link distances \( Z_A \) and \( Z_B \) between the entangled photon source and the two receiver nodes. The results in Fig.~\ref{ch1} show that the secret key rate rapidly decreases as the link distances increase. Similarly, as shown in Fig.~\ref{ch2}, the QBER grows with longer link distances due to the increased geometric loss and pointing error variance.
Overall, both figures highlight that at longer propagation distances, neither the key rate nor the QBER can satisfy the operational thresholds required for secure quantum communication. This limitation is particularly important for dynamically reconfigurable satellite constellations where maintaining consistent link performance over time is a major challenge.

\begin{figure}
	\centering
	\subfloat[] {\includegraphics[width=3.35 in]{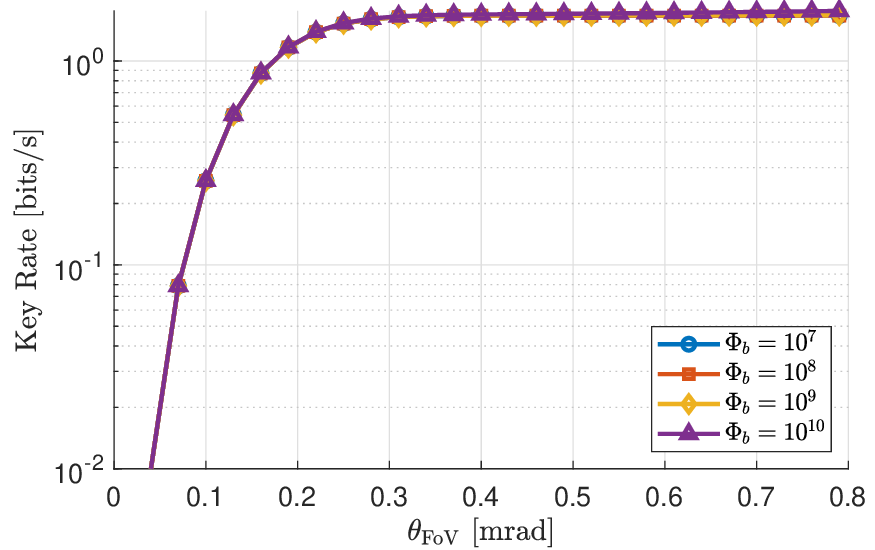}
		\label{ck1}
	}
	\hfill
	\subfloat[] {\includegraphics[width=3.35 in]{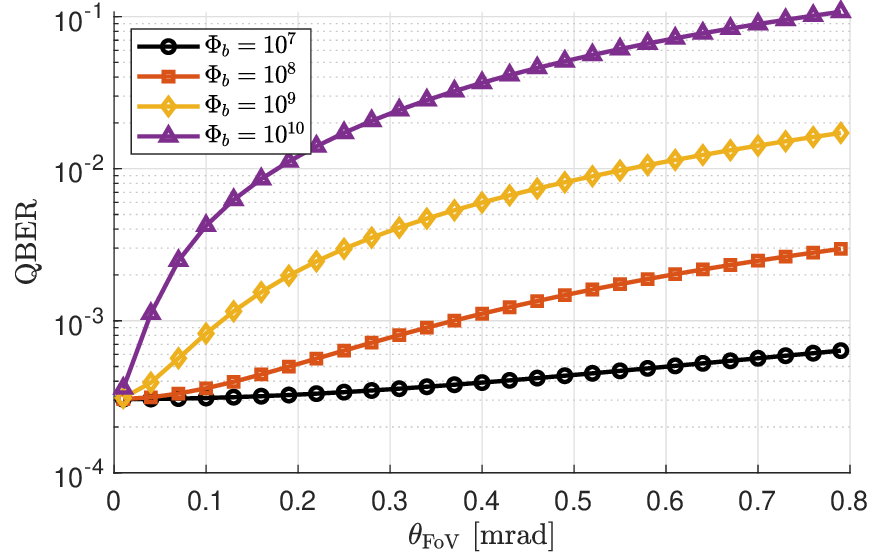}
		\label{ck2}
	}
	\caption{
		Effect of receiver FoV angle and background photon flux \(\Phi_b\) on system performance. (a) Average secret key rate versus FoV angle for various \(\Phi_b\) values, assuming symmetric link lengths of \(500\,\mathrm{km}\). (b) Corresponding QBER values. Increasing FoV enhances photon collection but also raises vulnerability to background noise, especially at higher \(\Phi_b\) levels.
	}
	
	\label{ck}
\end{figure}
%

\begin{figure}
	\centering
	\subfloat[] {\includegraphics[width=3.35 in]{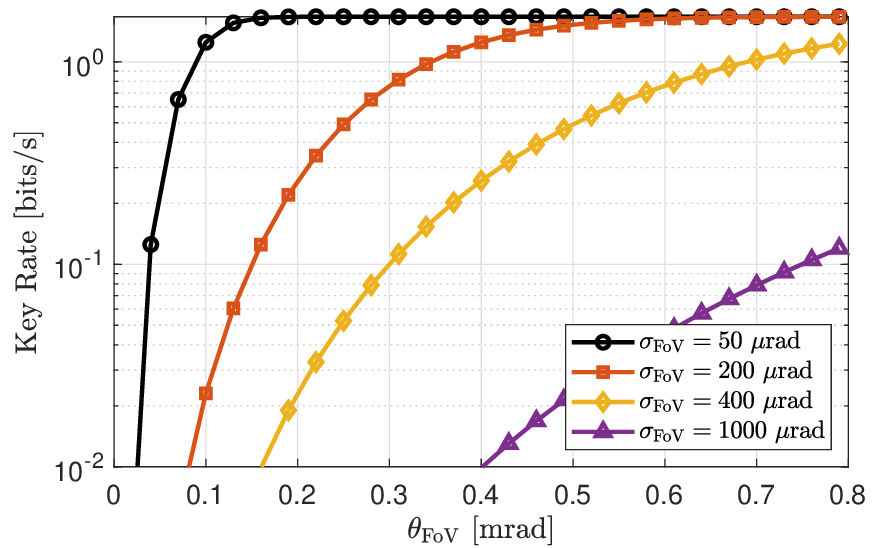}
		\label{cv1}
	}
	\hfill
	\subfloat[] {\includegraphics[width=3.35 in]{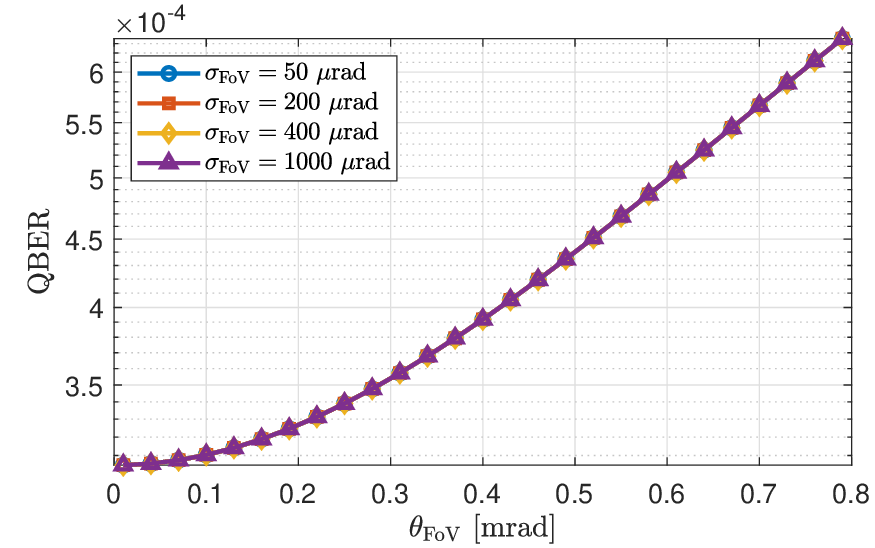}
		\label{cv2}
	}
\caption{(a) Secret key rate and (b) QBER as functions of the FoV angle for various levels of receiver angular jitter \( \sigma_{\mathrm{FoV}} \in \{50, 200, 1000\}~\mu\mathrm{rad} \). As \( \sigma_{\mathrm{FoV}} \) increases, more signal photons fall outside the detection cone and are lost, significantly reducing the key rate. Meanwhile, QBER remains almost unaffected since background photons are not filtered by angle.}	
	\label{cv}
\end{figure}
%

To further investigate the impact of tracking error on quantum communication performance, Fig.~\ref{cb1} and Fig.~\ref{cb2} present the average key generation rate and QBER, respectively, for different values of transmitter angular misalignment, \( \sigma_{\theta,t} \in \{1, 3, 5, 10\} \, \mu\mathrm{rad} \). The results are plotted for a fixed Bob link distance of \(500\,\mathrm{km}\), while varying Alice's link distance from \(100\) to \(1000\,\mathrm{km}\). 
As shown in Fig.~\ref{cb1}, reducing the tracking precision from \(1\,\mu\mathrm{rad}\) to \(10\,\mu\mathrm{rad}\) significantly degrades the key rate, especially at longer distances. This is due to the compounded beam divergence at higher propagation ranges, which reduces the probability of photons reaching the receiver aperture within the expected time slot. Since a valid quantum key bit is only formed when both Alice and Bob detect a photon within the same time window, the simultaneous degradation on both links leads to a sharp decline in key generation performance.
In parallel, Fig.~\ref{cb2} shows that both increasing the link distance and higher pointing jitter lead to a significant increase in QBER, often exceeding the acceptable threshold of \(10^{-2}\) required for reliable quantum key generation.

Another key parameter that directly influences the secret key rate is the photon pair generation rate \(\mu_t\). In the entanglement-based protocols, it is crucial to keep \(\mu_t\) sufficiently small to avoid generating multiple entangled photon pairs per time slot, as this may lead to ambiguity regarding the photon pair origin and thus introduce logical errors. However, in long-distance satellite quantum communication scenarios where transmittance losses are significant, increasing \(\mu_t\) can partially compensate for these losses by boosting the likelihood of detecting at least one photon at each receiver.
To better understand this trade-off, Fig.~\ref{cn1} and Fig.~\ref{cn2} depict the average key generation rate and QBER, respectively, for a range of \(\mu_t\) values from \(0.05\) to \(0.3\). As expected, higher \(\mu_t\) improves the key rate significantly (Fig.~\ref{cn1}) by increasing the overall signal photon flux. Nonetheless, this comes at the cost of elevated QBER (Fig.~\ref{cn2}) due to increased multi-pair emissions. These results underscore the importance of selecting \(\mu_t\) carefully to meet system-specific performance targets, balancing between acceptable error rates and desired throughput.

To isolate the effect of background photon flux on system performance, Fig.~\ref{ck1} and Fig.~\ref{ck2} present the average secret key rate and QBER, respectively, for a range of \(\Phi_b\) values under varying receiver FoV angles. The simulation considers symmetric link lengths of \(500\,\mathrm{km}\) for both Alice and Bob.
It is important to note that \(\Phi_b\) is a time-varying parameter influenced by environmental conditions, such as the presence of stars or celestial objects in the background. As expected, Fig.~\ref{ck2} shows that increasing \(\Phi_b\) results in a clear degradation in QBER performance due to a higher probability of detecting background photons instead of genuine entangled photons. Furthermore, both figures demonstrate that increasing the FoV angle leads to a higher photon detection probability, improving the secret key rate in Fig.~\ref{ck1}, but at the cost of increased QBER in Fig.~\ref{ck2}. This trade-off necessitates careful optimization of the FoV to balance throughput and key integrity.
Interestingly, while \(\Phi_b\) significantly degrades QBER, its effect on the key rate remains marginal, as shown in Fig.~\ref{ck1}.

Finally, we evaluate the impact of the receiver's angular fluctuation on the system performance. As discussed earlier, in entanglement-based QKD systems with narrow FoV filters, only photons whose incident angles lie within the FoV cone are detected. This makes the system highly sensitive to angular deviations at the receiver. In Fig.~\ref{cv1} and Fig.~\ref{cv2}, the secret key rate and QBER are plotted, respectively, versus the FoV angle \( \theta_{\mathrm{FoV}} \), for various values of the angular jitter standard deviation \( \sigma_{\mathrm{FoV}} \in \{50, 200, 1000\}~\mu\mathrm{rad} \). 
As shown in Fig.~\ref{cv1}, increasing \( \sigma_{\mathrm{FoV}} \) leads to a significant drop in the secret key rate. This is because a higher angular fluctuation increases the probability that received signal photons fall outside the FoV, and hence, are discarded by the optical filter. On the other hand, as evident from Fig.~\ref{cv2}, the value of \( \sigma_{\mathrm{FoV}} \) has negligible effect on the QBER. This is due to the fact that background photons are uniformly distributed over all angles and are not subject to the same directional coupling constraints as signal photons. Therefore, receiver-side angular jitter mainly reduces the number of useful signal detections without affecting the noise level, thereby degrading the overall system efficiency.

\section{Conclusion}
This paper investigated the fundamental performance limits of entanglement-based QKD over long-distance satellite-to-satellite links using the entanglement-based protocols. A detailed physical-layer model was developed that incorporates the effects of beam divergence, transmitter tracking error, receiver misalignment, narrow FoV, background noise, and multi-pair photon emissions. Analytical expressions were derived for the signal photon detection probability, background-induced error contributions, and the QBER, enabling accurate assessment of system performance under practical space-based conditions.

Our results reveal that the secure key generation rate is highly sensitive to link distance and angular misalignment, with severe degradation at longer ranges or under poor tracking conditions. Moreover, background photon interference—especially in narrow-FoV receivers—can significantly elevate QBER, posing a key challenge for reliable key generation. We further demonstrated that proper optimization of the photon pair generation rate, receiver FoV, and tracking precision is critical for maximizing secret key throughput while keeping the QBER below security thresholds.
The proposed framework offers valuable insights for the design and deployment of future satellite-based quantum communication networks. Future work may extend this model to dynamic orbital scenarios, adaptive tracking mechanisms, and integration with satellite-based quantum repeater protocols for global-scale entanglement distribution. Additionally, further analysis on other (more secure) entanglement-based protocols, such as device-independent QKD protocols, would provide even deeper insights into the effects of the physical layer on entanglement-based QKD protocols.

\balance

\end{document}